\documentclass{revtex4-2}
\usepackage[utf8]{inputenc}
\usepackage{color}
\usepackage[margin=1.0in]{geometry}
\usepackage{amsmath}
\usepackage{amssymb}
\usepackage{amsmath}
 \usepackage{appendix}
\usepackage{bbold}
\usepackage{setspace}

\begin{document}

\title{Time Fisher Information associated with Fluctuations in Quantum Geometry}

\author{Salman Sajad Wani}
\affiliation{Canadian Quantum Research Center 204-3002, 32 Ave Vernon, BC V1T 2L7 Canada}

%\affiliation{School of Physics, Damghan University, P. O. Box 3671641167, Damghan, Iran}

\author{James Q. Quach}
\affiliation{Institute for Photonics and Advanced Sensing and School of  Physical Sciences, The University of Adelaide, South Australia 5005, Australia}

\author{Mir Faizal}
\affiliation{Canadian Quantum Research Center 204-3002, 32 Ave Vernon, BC V1T 2L7 Canada}
\affiliation{Irving K. Barber School of Arts and Sciences, University of British Columbia - Okanagan, Kelowna, British Columbia V1V 1V7, Canada }
\affiliation{Department of Physics and Astronomy, University of Lethbridge, Lethbridge, AB T1K 3M4, Canada}

\begin{abstract}
As time is not an observable, we  use  Fisher information (FI) to address the problem of time. 
We  show that the Hamiltonian constraint operator cannot be used to analyze any quantum process for  quantum geometries that are associated with time-reparametrization invariant classical geometries. This is because the Hamiltonian constraint does not contain FI  about time. We demonstrate that although the Hamiltonian operator is the generator of time, the Hamiltonian constraint operator can not observe the change that arises through the passage of time. This means that the problem of time is inescapably problematic in the associated quantum gravitational theories. Although we explicitly derive these results on the world-sheet of bosonic strings, we argue that it  holds in general. We also identify an operator on the world-sheet which contains  FI about time in a string theoretical processes. Motivated by this observation, we propose that a criteria for a meaningful operator of any quantum gravitational process, is that it should contain non-vanishing FI  about time. 

%We relate the problem of time to the  inability of the Hamiltonian constraint to probe any change, and this is in turn related to the absence of quantum Fisher information (FI) of the  Hamiltonian constraint  with respect to  quantum gravitational   states. Even though we explicitly derive our results for the world-sheet of bosonic string theory, this argument    holds for any quantum geometry associated with time-reparametrization invariant classical geometry. We also identify an operator on the world-sheet  which contains FI with respect to string states, and we explicitly demonstrate that it can be used to probe changes in a string theoretical processes. Motivated by this observation, we propose that a criteria for a meaningful operator in any quantum gravitational process is that it should contain non-vanishing FI with respect to  quantum gravitational states.
\end{abstract}

\maketitle

It is known that for any  classical  geometry with  time-reparametrization invariance, the Hamiltonian represents the  temporal part of the diffeomorphism constraint    \cite{adm, adm1, adm2, adm4}. The   Hamiltonian  constraint  is  a  generator of gauge transformation, due to the diffeomorphism invariance being a gauge degree of freedom. This leads to an absence of  physical time in quantum gravity, and this absence of time is called the problem of time  ~\cite{pt12, pt14}. This problem occurs in almost all approaches to quantum gravity, such as the  Wheeler-DeWitt approach \cite{wh, wh0}, loop quantum gravity \cite{wh2, wh21}, discrete quantum gravity \cite{wh4, wh41}, group field theory \cite{wh5, wh51}, quantized modified   gravity \cite{wh6, wh61}, or in the quantization of both brane world theories  \cite{wh7, wh71} and Kaluza–Klein geometries  \cite{wh8, wh81}.  It has been suggested that the problem of time can be resolved by novel interpretations of quantum gravity, such as in the frozen formalism \cite{pt12}, the use of suitable operators for Cauchy surfaces \cite{pt14}, the use of matter fields as time  \cite{ab10}, the use of scale factors as time \cite{ab15}, the matrix formulation of quantum gravity \cite{ab12},  unimodular gravity in the  Ashtekar formulations \cite{ab14}, the rigging map of group averaging \cite{ab16}, conditional probabilities \cite{ab18},  non-perturbative quantum gravity \cite{ab19, ab20}, third quantization \cite{ab21, ab22} and even  fourth quantization \cite{ab24}. 

% Time is not an observable in quantum mechanics \cite{time12}, and such    quantities which are not associated with an observable can be investigated using Fisher information (FI) \cite{eq1, eq2, eq4, eq5, eq6, eq7}. 
Part of the enigma of time is that it is not an observable in quantum mechanics \cite{time12}. This motivates us to address the problem of time through an information-theoretic lens.  Specifically, we consider time to be a hidden or unobservable variable which one may only indirectly probe through quantum operators. The amount of information that one may extract about unobservales through  observables is given by the FI \cite{eq1, eq2, eq4, eq5, eq6, eq7}. To analyze the FI of time, we first observe that time is associated with changes in all the aforementioned proposed solutions to the problem of time.  
The quantity associated with changes in Hamiltonian is work. At the microscopic scale however, work is a notoriously subtle concept as quantum fluctuations are on the same order of magnitude as expectations values \cite{campisi11,jarzynski11,talkner07}. As such there is no single definition of work distributions in quantum theory, and several schemes exists \cite{baumer18}. The two-point measurement (TPM) scheme is the most established, where the work distributions is obtained with two projective measurements of the system energy at the beginning and end of a process \cite{tasaki00,kurchan01}. The TPM however can not be applied to relativistic systems, such as in quantum field theory (QFT), as the projective measurements may lead to locality violation and superluminal signals \cite{5cd, 5mn, 6lk}. For this reason, the Ramsey scheme was developed to construct the work distribution for relativistic systems \cite{5, 6ab, 6ba}. Here an auxiliary qubit is coupled to the system, and information about the system is transferred to the qubit, where measurements are done. Specifically, the qubit engages the system in an evolution conditional on whether the qubit is excited or not. By preparing the qubit in a superposition of ground and excited states, this process transfers the data about the characteristic function of the TPM work distribution to the state of the qubit. This non-invasive procedure acquires statistics which otherwise would require projective measurements. It has been shown that the work distribution obtained with the Ramsey scheme is well defined for QFT, even though project measurements may not be \cite{5}.

The notion that energetic change or work can reveal something about time can be formalised with the FI  \cite{eq1, eq2, eq4, eq5, eq6, eq7}. This is because the FI quantifies the amount of information that an observable random variable, in this case the work, provides about an unknown or hidden parameter, in this case time. In this letter, we will extend the Ramsey interferometric scheme to the world-sheet of bosonic strings, to show the impossibility of the Hamiltonian constraint operator to probe change and therefore time. We will show that the associated work distribution contains no FI with respects to time. We offer the mass-squared operator as an alternative operator that may probe time with, non-zero time FI.

%In this letter, we will extend the Ramsey interferometric scheme to the world-sheet of bosonic strings, to show the impossibility of the Hamiltonian constraint operator to probe change and therefore time. The reason for this is that the work distribution contains no Fisher information (FI)  about time. In this context we   take a deeper look at quantum work, which is basically represented by the difference between eigenvalues of the Hamiltonian. We will investigate the possibility of probing the change in the quantum states, by performing such an analysis for an  alternative operator  (mass-squared operator). We will be able to show that this alternative mass-squared operator     contains FI about time. This will be  done by  deriving  an  expression analogous to  the work distribution,  but this time for the mass-squared operator,  and demonstrating that unlike the Hamiltonian constraint operator, the FI  does not vanish for it.
%It may be noted that in both these proposals, we have calculated the FI about time, as FI can be used to obtain information about quantities which are not directly represented by quantum observables. As time is such a quantity, it is important to calculate the FI about time form the distribution of the difference of the eigenvalues of specific operators during a quantum process (which exists for a specific time interval). 

\label{sec:Hamiltonian constraint operator} \emph{Hamiltonian constraint operator.} For  open bosonic strings with Neumann boundary conditions,  we can expand the operator corresponding to the  world-sheet fields  $ \hat{X}^\mu (\tau,\sigma)$ in terms of modes as 
\begin{equation} 
    \hat{X}^\mu (\tau,\sigma)= \hat{x}^\mu  +2\alpha' \tau \hat{p}^\mu +i \sqrt{2\alpha'}\sum_{n\neq 0}\frac{1}{n}\hat{\alpha}^{\mu }_n e^{-i  n\tau}\cos n\sigma~,
    \end{equation}
where  $ \hat{x}^\mu $ is the center of mass, $\hat{p}^\mu $ is  the momentum of the center of mass,  $\hat{\alpha}^{\mu }_n$ are the string oscillatory modes, and $\alpha'$ is the string length scale. Here $\sigma \in [0, \pi) $ and $\tau$ are the spatial and temporal world-sheet coordinates, respectively. 
 
We explicitly introduce change into the system by perturbing it with a dilaton field $\phi(X)$, which couples to the  world-sheet curvature $R$ in the standard way. Expanding  the dilaton field   as $\phi(X) \approx \phi_0 + (\partial_\mu \phi) \hat{X}^{\mu } $,    setting $\phi_0=0$, and absorbing the time dependence in $\chi (\tau) $, we can write  $ \phi(X) \approx  \lambda\chi (\tau) c_\mu \hat{X}^{\mu} $, where $c_\mu $ is a constant vector.  Now for an interacting Hamiltonian,   $\chi(\tau)$ will act as a  switching function which turns on the interaction for finite duration, $\lambda$ will be viewed as a coupling constant.  Thus,  the Hamiltonian of the perturbed string can be written as 
  \begin{equation}
    \hat{H}_X(\tau) \approx  \hat{H} +  \lambda\chi (\tau) \int d\sigma R c_\mu  \hat{X}^\mu  =\hat{{H}}+\hat{{H}}_I(\tau) 
    \end{equation}
with $\hat{H}$ as the free string Hamiltonian. The interaction with the dilaton field is chosen such that perturbation only exists between $\tau >0$ and $\tau <t $. The  unitary evolution operator resulting from the dilaton field perturbation is  
    \begin{equation}
      \hat{U} (t) = \mathcal{T} \exp{\left(-i \lambda\int_0^t d\tau \hat{{H}}_I(\tau) \right)}, 
    \end{equation}
where $\mathcal{T}$ denotes time ordering between $0<\tau<t$.

The change in energy of the string is characterised by the work probability distribution with distribution variable $\mathcal{H}$   \cite{Za, 5ab, 5ba} 
\begin{equation}
	P(\mathcal{H})=\sum_{il}p_{il}\delta(\mathcal{H}-\Delta H_{il}) 
\end{equation}
with the possible values of work $ \Delta{H}_{il}\equiv E'_l-E_i$  being the difference between the initial and final eigenvalues of $\hat{H}(\tau)$, and $p_{il}= \langle H_i|\hat{\rho}|H_i\rangle{\langle H'_l|U|H_l\rangle}^2$, where $|H_l\rangle$ are the initial eigenstates and $|H'_l\rangle$ the final eigenstates of  the free string Hamiltonian. Associated with the probability distribution is the characteristic function of a 
 real-valued variable  $\theta$, expressed as 
\begin{equation}
	\Tilde{P}(\theta) =\int d \mathcal{H} P(\mathcal{H}) e^{i \theta  \mathcal{H}} =\langle e^{i \theta  \mathcal{H}} \rangle~
\label{ba}
\end{equation}

In the TPM scheme, the work probability distribution is obtained with projective measurements of $E_i$ and $E_l'$. As discussed above this is problematic for systems with a Lorentz structure, as it is incompatible with relativistic causality. To overcome this problem we follow the Ramsey interferometric scheme, and   prepare a pure string state $\sum_n d_n |n;p\rangle$ (with  $n$ representing the string oscillatory modes, and $p$  representing the momentum of the center of mass), and then using a combination of $n,m$ different string   oscillatory states to write: $ \hat{\rho}= \mathcal{N} \sum_l\sum_k d_k{d_l}^*|k;p\rangle\langle l;p|$.   
We couple an auxiliary qubit to the string. The string and qubit (which is initially in the ground state) are prepared in a product state, $\hat{\rho}_\text{tot}=   \hat{\rho}\otimes\hat{\rho}_\text{aux}$. A Hadamard operator is then applied to the qubit. The state of the total system is then dictated by the unitary evolution operator \cite{5} 
 \begin{equation}
     \hat{C}_\theta  (t) = \hat{U} e^{-i\theta  \hat{H}(0) } \otimes|0\rangle\langle0|+ e^{-i\theta  \hat{H}(t)}\hat{U}\otimes|1\rangle\langle1|~,
 \end{equation}
with the qubit state given by $\hat{\rho}_\text{aux}= \text{Tr} _{X}[\hat{C}_{\theta}(t)\hat{\rho}_\text{tot} \hat{C}^\dagger_{\theta}(t)], 
$ where  $\text{Tr}_{X}$ is a trace over the string states. Applying a final Hadamard operation, the qubit state in the weak coupling limit is (see Appendix 1) 
\begin{equation}
    \hat{\rho}_\text{aux}= \frac12 (\mathbb{1} +\hat{\sigma}_z)~.   
\label{eq:rho_aux_string}
\end{equation}
We make the observation that in general  \cite{6ab, 6ba}  
\begin{equation}
\hat{\rho}_\text{aux}    =
\frac12\Big\{\mathbb{1} + \text{Re} [\tilde{P}(\theta )]\hat{\sigma}_z + \text{Im} [\tilde{P}(\theta )]\hat{\sigma}_y\Big\}~. 
\label{eq:rho_aux}
\end{equation}

Comparing Eq.~(\ref{eq:rho_aux_string}) with Eq.~(\ref{eq:rho_aux}), we see that the characteristic function for the bosonic string is
    \begin{equation}
    \tilde{P}(\theta )=1~.
\label{za}
 \end{equation} 
Taking the  first moment of Eq. (\ref{ba}), the  average difference between the initial and final eigenvalues of $\hat{H}_X$ (work) is 
\begin{eqnarray}
 \langle \mathcal{H} \rangle 
 = -i \frac{d}{d\theta }  \tilde{P}(\theta )|_{\theta =0} = 0~.
\label{eq:H_av}
\end{eqnarray} 
The absence of work occurs due to the inability of the Hamiltonian constraint to have any    information about time. This is directly observed through the FI of the work distribution with respects to time \cite{eq1, eq2, eq4, eq5}  
\begin{equation}
    F(t)=\int P(\mathcal{H})|\frac{\partial }{\partial t}\log P(\mathcal{H})|^2d\mathcal{H}=0~,
\end{equation}
where $ P(\mathcal{H}) =\delta(\mathcal{H})$ is the inverse Fourier transform of Eq.~(\ref{za}). Even though we had explicitly introduced change by perturbing the system, and indeed the string states have changed as dictated by $\hat{U}$,  Eq.~(\ref{eq:H_av}) tells us that  $\hat{H}$ can have no 
information of this change.  \emph{In other words, we have made the seemingly paradoxical observation that even though the Hamiltonian can evolve string states, it can not be used to observe such evolution. }   

%{\color{blue}{ In fact,  for quantum work distribution, $ \langle \mathcal{H}^2 \rangle -  \langle \mathcal{H} \rangle^2 $ also vanishes. %We know that for for pure states, the FI  with respects to an operator $\hat{A} $  can be simplified to $F[\hat{\rho},\hat{A} ] = 4[\langle \hat{A}^2\rangle- \langle \hat{A}\rangle^2]$ \cite{eq4}. Thus, it seems natural to use FI to investigate this process.   However, it is not possible to define  an operator corresponding to work, and so we cannot directly obtain  FI for work distribution. We  argue that to measure change in states of a system, we need to first obtain information about those states  by operating on those states with an operator which contains FI. Thus, inability to measure quantum work in  string theoretical process  thus can be  related to the absence of FI in $\hat{H}$ with respect to the string states. We can observe that indeed that FI of $\hat{H}$ vanishes, due to the being   Hamiltonian is a constraint, with  
%$ \langle \hat{H}^2 \rangle = \langle \hat{H}^2 \rangle $, and so
 %\begin{equation}
  %   F[\hat{\rho},\hat{H} ] =0
 %\end{equation}
Even though this  result was derived explicitly  for string theory,  it holds not only for string theory but  any quantum geometry associated with a time-reparametrization invariant classical geometry, as the Hamiltonian is also a constraint in them.  In contrast, the FI of time  with respects to the Hamiltonian of a quantum field theory does not vanish, due to  the  non-vanishing  quantum work distribution of quantum fields \cite{5}. 

From another viewpoint, the FI is related to the Mandelstam-Tamm bound $\tau\ge 1/F$. As the Mandelstam-Tamm provides the quantum speed limit of evolution between states, vanishing FI with resepcts to time, suggests either a static universe or that the Hamiltonian operator cannot probe time. We will show that is  the latter, as we find an alternative operator that yields a finite Mandelstam-Tamm bound. 
   
\label{sec:Mass-squared operator} \emph{Mass-squared operator.} If one considers the mass-energy dispersion relation in quantum field theory, the mass of a field is constant, whilst its energy can vary as it interacts with its environment. In contrast, the energy of a string is a constant, as the Hamiltonian in string theory is a constraint, and it is its mass that can be treated as a dynamical variable. As  it is natural to classify states using string oscillatory modes, and the  information about string oscillations is captured by the operator representing mass-squared $\hat{M}^2$, we    identify   $\hat{M}^2$  as alternative operator to probe the change in  strings states. This operator can be explicitly  expressed in terms of the string oscillatory modes as 
 \begin{equation}
\hat{M}^2 = \frac{1}{\alpha'}\Bigg(\sum_{i=1}^{24}\sum_{n>0}
\hat{\alpha}_{-n}^i\hat{\alpha}^i_n-1\Bigg)~.
 \end{equation}

Now following what was done for the Hamiltonian operator, we define a  distribution variable for  $\hat{M}^2$ as  $\mathcal{M}^2 $, such that    $P(\mathcal{M}^2)=\sum_{il}p_{il}\delta(\mathcal{M}^2-\Delta M_{il}^2)$, where  $p_{il}= \langle M^2_i|\hat{\rho}|M^2_i\rangle{\langle M'^2_l|U|M^2_l\rangle}^2$ is  the associated joint probability distribution, and $\Delta{M^2}_{il}$ is the difference between the  final   and initial   eigenvalues of $\hat{M}^2$.  Thus,  we can write the corresponding characteristic function for $\mathcal{M}^2 $ as 
 \begin{equation}
 \Tilde{P}(\theta ) =\int d \mathcal{M}^2 P(\mathcal{M}^2) e^{i \theta  \mathcal{M}^2} =\langle e^{i \theta  \mathcal{M}^2} \rangle
 \end{equation}

Applying the Ramsey scheme, the unitary evolution operator is 
 \begin{equation}
	\hat{C}_{\theta}(t) = \hat{U} e^{-i\theta  \hat{M^2}(0) } \otimes|0\rangle\langle0|+ e^{-i\theta  \hat{M^2}(t)}\hat{U}\otimes|1\rangle\langle1|~.
\label{eq:C_M2}
\end{equation}
 Here the systems evolution is governed by $\hat{{H}}_I$, but importantly, the changes in states are probed with $\hat{M}^2$. Following the previous analysis (see  Appendix 2), we observe that  Eq.~(\ref{eq:C_M2}) leads to the characteristic function
\begin{equation}
  \begin{split}
  \Tilde{P}(\theta )=&1 +\lambda^2\mathcal{N} {\alpha}'\sum^\infty_{n=1}\sum^\infty_{k=0}\frac{d^2_k}{n^2}(n+2k)\Tilde{\chi}(n)\Tilde{\chi}(-n)|\Phi(n)|^2   \left[\exp{\left(i \frac{\theta  n}{\alpha'}\right)} -1 \right]~, 
  \end{split}
  \label{eq:PM}
\end{equation}
    where $\Tilde{\chi}(n)     $ 
    is a Fourier transformation  of  the switching function $\chi(\tau)$ and $|\Phi(n)|^2 = \Phi(n)^{\mu}\Phi(n)_{\mu} $, with $\Phi(n)_{\mu }$  as  the  cosine transformation  of the $Rc_\mu $ (for constant $c_\mu $). 
Taking the first moment of    $ \tilde{P}(\theta )=\langle e^{i \theta   \mathcal{M}^2} \rangle $, the  average difference between the initial and final eigenvalues of  $\hat{M}^2$  is   
\begin{equation}
\langle  \mathcal{M}^2\rangle =\mathcal{N}\lambda^2 \sum^\infty_{n=1}\sum_{k=0}^\infty \frac{d^2_k}{n} (n+2k)|\Tilde{\chi}(n)|^2|\Phi(n)|^2
\end{equation}
As this does not vanish in general, one can analyze the changes in string states with  $\hat{M}^2$. 

Taking the inverse Fourier transform of Eq.~(\ref{eq:PM}),
\begin{eqnarray}
    P(\mathcal{M}^2)=\delta(\mathcal{M}^2)+\lambda^2\mathcal{N}\alpha'\sum^\infty_{n=1}\sum^\infty_{k=0}(2+2\cos nt)&&\nonumber\\|\hat{\Phi(n)}|^2\frac{d^2_k(k+n)}{n^3} \left[\delta\left(\frac{n}{\alpha'}-\mathcal{M}^2
    \right)-\delta(\mathcal{M}^2)\right]&& \nonumber~,   \\  
\end{eqnarray}
we can write down the FI of work distribution of the $\hat{M}^2$ with respects to time \cite{eq1, eq2, eq4, eq5}  
\begin{eqnarray}
   F(t)=\int P(\mathcal{M}^2)|\frac{\partial }{\partial t}\log P(\mathcal{M}^2)|^2d\mathcal{M}^2~.
\end{eqnarray}
This integral can be expressed as 
\begin{eqnarray}
    F(t)%=\frac{S^2}{1-K}+(\lambda^2\mathcal{N}\alpha')^2\frac{\sum_k\sum_{k'}\sum_{n\neq0}\frac{d_k^2d_{k'}^2(n+2k)(n+k')}{n^2}sin^2(nt)|\phi(n)|^4}{K}\nonumber\\
    =\frac{\Dot{K}^2}{1-K}+(\lambda^2\mathcal{N}\alpha')^2
    \frac{Y}{K}
\end{eqnarray}
where $\Dot{K}$ is the time derivative of $K$, with $    K=\lambda^2\mathcal{N}\alpha'\sum^\infty_{n=1}\sum^\infty_{k=0}{d_k^2(n+2k)}{n^{-3}}[2+2 \cos nt]|\Phi(n)|^2, 
   $ and $
    Y = \sum^\infty_{n=1}\sum^\infty_{k=0}\sum^\infty_{k'=0}  {d_k^2d_{k'}^2(n+2k)(n+2k')} {n^{-2}}  \sin^2 nt|\Phi(n)|^4  $. 
%\begin{eqnarray}
  %  S&=&\lambda^2\mathcal{N}\alpha'\sum_k\sum_n\frac{d_k^2(n+2k)}{n^2}2\sin(nt)|\Phi(n)|^2\\ 
   % K&=&\lambda^2\mathcal{N}\alpha'\sum^\infty_{n=1}\sum^\infty_{k=0}\frac{d_k^2(n+2k)}{n^3}[2+2 \cos nt]|\Phi(n)|^2  
    %\\ 
    %Y &=& \sum^\infty_{n=1}\sum^\infty_{k=0}\sum^\infty_{k'=0}\frac{d_k^2d_{k'}^2(n+2k)(n+2k')}{n^2}\sin^2 nt|\Phi(n)|^4  \nonumber\\
%\end{eqnarray}
As in general $|\Phi(n)|^2 \neq 0$ for a dilaton field, the FI of $\hat{M}^2$ with respects to time does not vanish, in stark constrast to $\hat{H}$. In fact, it is exactly for this reason that $\hat{M}^2$  has been  used to obtain non-trivial results  about  string states rather than $\hat{H}$ \cite{t1, t2}, even though it has never been explicitly explained from this information-theoretic viewpoint. Motivated  by this observation, we propose that  an essential property for an operator to analyze change in a  quantum gravitational process,  is that its   FI with respects to time should not vanish from the distribution of the difference between eigenvalues of  that operator, $F(t) \neq 0$.

\label{Conclusion}\emph{Outlook.}
We demonstrated that the inability of the Hamiltonian constraint operator to probe change is related to the absence of FI with respect to time. We argue this holds for any quantum geometry associated with time-reparametrization invariant classical geometry. From this one may conclude that either the Hamiltonian constraint operator is an inappropriate operator for the probing of time, or  that time-reparametrization invariance should be broken for time to exists. We showed that the mass-squared operator is a natural alternative that may probe time. That time-reparameterization invariance should be broken certainly is also an avenue worth pursuing, and our work provides the information-theoretic framework to consider the problem of time in universes with spontaneously broken time-reparametrization invariance. 

\section*{Acknowledgements} 
We would like to thank D. J.  Smith for suggesting that we perturb the string with a dilaton field. We would also like to thank  A.~Teixid\'o-Bonfill,  E.~Mart\'\i{}n-Mart\'\i{}nez and A.~Ortega, for discussion on quantum work distribution. J.Q.Q. acknowledges the Ramsay fellowship for ﬁnancial support of this work.

\section*{Appendix 1. Hamiltonian Constraint Operator}
The  open bosonic strings with Neumann boundary conditions,  can be expanded   in terms of modes as 
\begin{equation} 
    \hat{X}^\mu  (\tau,\sigma)= \hat{x}^\mu   +2\alpha' \tau \hat{p}^\mu  +i \sqrt{2\alpha'}\sum_{n\neq 0}\frac{1}{n}\hat{\alpha}^{\mu }_n e^{-i  n\tau}\cos n\sigma~,
    \end{equation}
where  $ \hat{x}^\mu  $ is the center of mass, $\hat{p}^\mu  $ is  the momentum of the center of mass,  $\hat{\alpha}^{\mu }_n$ are the string oscillatory modes, and $\alpha'$ is the string length scale.  
We start by   preparing  an  initial string-qubit system as
 $\sum_k d_k |k;p\rangle$ (with  $k$ representing the string oscillatory modes, with  $p$ as the momentum of the center of mass). Using  a combination of $k,l$ different string   oscillatory modes, we write   $ \hat{\rho}= \mathcal{N} \sum_k\sum_l  d_k{d_l}^*|k;p\rangle\langle l;p|$, where $ \mathcal{N}=1/\sum_k |d_k|^2$. %Here $|k;p\rangle$ is the  $k$ oscillatory state of  strings, with fixed momentum $p$, and  $d_k$ is its associated coefficient.  
 We couple this string state  to an 
  auxiliary qubit,  and define  $\hat{\rho}_\text{tot}=   \hat{\rho}\otimes\hat{\rho}_\text{aux} = \hat{\rho}\otimes|0\rangle\langle0|$, where   $|0\rangle\langle0|$ is the ground state of the string auxiliary qubit. 
%\jam{define separable state as per main text above Eq. 6. Also change all $\rho_b$ to $\rho_\text{aux}$}
%\begin{eqnarray}
  %    \hat{\rho}_\text{tot}=\hat{\rho}\otimes|0\rangle\langle0|, &&
   %     \hat{\rho}=  \mathcal{N} \sum_k\sum_l\ d_kd^*_l|k;p\rangle\langle l;p|
  %  \end{eqnarray}
 Upon applying the first Hadamard gate on this string auxiliary qubit, we obtain  $
 \hat{\rho}_\text{tot}=\hat{\rho} \otimes|+\rangle\langle+|~
$. Now we perturb it by a dilaton field. 
 The Hamiltonian for the  system perturbed by the dilaton field can be written as 
    \begin{equation}
    \hat{H}_X(\tau)=\hat{{H}} +  \lambda\chi (\tau) \int d\sigma R c_\mu  \hat{X}^\mu   =\hat{{H}} +\hat{\Tilde{H}}_I(\tau) 
    \end{equation}
Here  we have expanded the dilaton field as  
    \begin{equation}\phi(X) \approx \phi_0 + (\partial_\mu \phi) \hat{X}^\mu   = \lambda\chi (\tau) c_\mu  \hat{X}^\mu  ,  \label{b1}
    \end{equation} 
    where $\phi_0 =0$,   $\hat{H}$  is the free Hamiltonian of the system,   
$R$ is the scalar curvature on the world-sheet, and $c_\mu $ is  a constant vector.    We have taken the standard coupling of the strings states to a dilaton field.
  Here the switching function is  $\chi(\tau) $ and $ R c_\mu  \hat{X}^\mu  $ is the  smearing function. This dilaton field $\phi(X)$ interacts with the strings, and this perturbation can be expressed using   a unitary $\hat{U}  (t) $, such that % such that   $\hat{\Tilde{H}}_X(0)= \hat{\Tilde{H}}_X(t)  $,  
    \begin{equation} \label{a}
      \hat{U} (t)  =\mathcal{T} exp{\left(-i \lambda\int d\tau \chi(\tau)\int d\sigma R c_\mu  \hat{X}^\mu  (\tau,\sigma))\right)}
    \end{equation}
    where   $\mathcal{T}$ denotes time ordering between $0<\tau<t$.
    Using the  Dyson expansion, we obtain 
    \begin{equation}
    \hat{U} (t)  =1+\hat{U}^{(1)}+\hat{U}^{(2)}+\hat{O}(\lambda^3)
    \end{equation}
    with 
    \begin{eqnarray}
        \hat{U}^{(1)}&=&-i \lambda\int d\tau \hat{H}_I(\tau)
   \\ 
        \hat{U}^{(2)}&=&-\lambda^2\int  d\tau\int  d\tau'\hat{H}_I(\tau)\hat{H}_I(\tau').
        \end{eqnarray}
The state of the total system is then evolved by the unitary  operator, and expressed using  $\hat{C}_\theta (t)$ (with $\theta$ as the real-valued variable used in the definition of the characteristic function for this system)  
    \begin{equation}
 \hat{C}_{\theta}(t) = \hat{U} (t)   e^{-i\theta   \hat{H}(0) } \otimes|0\rangle\langle0|+ \nonumber e^{-i\theta   \hat{H}(t)  }\hat{U} (t)  \otimes|1\rangle\langle1|
 \end{equation}
 The reduced states is a   qubit state, and can be written as   
 \begin{equation}
    \hat{\rho}_\text{aux}(t)= \text{Tr} _{X}[\hat{C}_\theta(t)\hat{\rho}_\text{tot} \hat{C}_\theta^\dagger (t)]
  \end{equation}  where $Tr_X$ is a trace over the $\hat{X}^\mu  $. 
Now  using the  Dyson expansion of the unitary operator $\hat{U}  (t)  $, we can write  
 \begin{equation}
{\hat{\rho}}_\text{aux} (t) ={\hat{\rho}}^{0}_\text{aux}(t)+{\hat{\rho}}^{1}_\text{aux}(t) + {\hat{\rho}}^{2}_\text{aux}(t)+\mathcal{O} (\lambda^3)
\end{equation}
Now we will  explicitly evaluate  these terms. The first order term in the Dyson expansion of the qubit state is
     \begin{equation}
        \hat{\rho}^0_\text{aux}(t)=|+\rangle\langle+|
     \end{equation}
We can also write  $\hat{\rho}^1(t)$ as  
     \begin{eqnarray}
   \hat{\rho}^1_\text{aux}(t)&=& 
   \text{Tr}_{X}\Big[\big( \hat{U}^{(1)} (t)   e^{-i\theta   \hat{H}(t)   } \otimes|0\rangle\langle0|
   \label{b4}  +  e^{-i\theta   \hat{H}(t)  }\hat{U}^{(1)} (t)  \otimes |1\rangle\langle1|\big) \hat{\rho}_\text{tot}\nonumber \\ && \times \big(e^{i\theta   \hat{H}(t)  } \otimes|0\rangle\langle0|+ e^{i\theta   \hat{H}(t)   }\otimes    |1\rangle\langle1|\big) \Big]\nonumber \\&&
    + \text{Tr}_{X}\Big[ \big(e^{-i\theta  \hat{H}(t)   } \otimes|0\rangle\langle0|+ e^{-i\theta   \hat{H}(t)  } \otimes |1\rangle\langle1|\big)\hat{\rho}_\text{tot}\nonumber\\&& \times 
   \big( e^{i\theta  \hat{H}(t)   }\hat{U}^{(1)\dagger} (t)   \otimes|0\rangle\langle0|+ \hat{U}^{(1)\dagger} (t)  e^{i\theta   \hat{H}(t)  } \otimes |1\rangle\langle1|\big)\Big]  
    \end{eqnarray} 
     Using the property $\hat{U}^{(1)} =-\hat{U}^{(1)\dagger} $, the coefficient of  $|0\rangle\langle0|$ is given by 
     \begin{eqnarray}
     \text{Tr}_X[\hat{U}^{(1)}  e^{-i\theta   \hat{H}(t)  }\hat{\rho}_\text{tot}e^{i\theta   \hat{H}(t)  }+e^{-i\theta   \hat{H}(t)  }\hat{\rho}_\text{tot}e^{i\theta   \hat{H}(t)  }\hat{U}^{(1)\dagger} ] =0 \end{eqnarray}
     It may be noted that by repeating such calculations for all the other coefficients of Eq. (\ref{b4}), we can observe that all those coefficients vanish,  and so we can write 
     \begin{equation}
   \hat{\rho}^1_\text{aux}(t) =0.       
     \end{equation}
    
Now we can write the  second order term as  
\begin{eqnarray}
   \hat{\rho}^2_\text{aux}(t)
  &=& \text{Tr}_X\Big[\big(\hat{U}^{(1)} (t)   e^{-i\theta   \hat{H}(t)  }\otimes|0\rangle\langle0|+ e^{-i\theta   \hat{H}(t)  }\hat{U}^{(1)} (t)   \otimes|1\rangle\langle1|\big)\nonumber
   \\&& \times\hat{\rho}_\text{tot}\big(e^{i\theta   \hat{H}(t)  } \hat{U} (t)  ^{(1)\dagger}\otimes|0\rangle\langle0|+ \hat{U} (t)  ^{(1)\dagger}e^{i\theta   \hat{H}(t)  } \otimes|1\rangle\langle1|\big) \Big] \nonumber
   \\&&
  + \text{Tr}_X\Big[\big( \hat{U}^{(2)} (t)   e^{-i\theta   \hat{H}(t)  }\otimes|0\rangle\langle0|  + e^{-i\theta   \hat{H}(t)  }\hat{U}^{(2)} (t)   \otimes|1\rangle\langle1|\big)\nonumber\\&& \hat{\rho}_\text{tot} \big(e^{i\theta   \hat{H}(t)  } \otimes|0\rangle\langle0|  + e^{i\theta   \hat{H}(t)  }\otimes|1\rangle\langle1| \big)\Big)  \nonumber\\&&
  + \text{Tr}_X\Big(\big(e^{-i\theta   \hat{H}(t)  }\otimes|0\rangle\langle0|+e^{-i\theta   \hat{H}(t)  } \otimes|1\rangle\langle1|\big)\hat{\rho}_\text{tot}\nonumber\\ && \times \big(e^{i\theta   \hat{H}(t)  }\hat{U}^{(2)\dagger} (t)  \otimes|0\rangle\langle0|+\hat{U}^{(2)\dagger} (t)  e^{i\theta   \hat{H}(t)  }\otimes |1\rangle\langle1| \big) \Big]
   \end{eqnarray}
This equation is of the form   
\begin{equation}
  \hat{\rho}^2_\text{aux}(t) = a_0 |0\rangle\langle0|+ a_1|1\rangle\langle1| + a_2 |0\rangle\langle1| + a_3 |1\rangle\langle0|
\end{equation}
To obtain $(a_0, a_1, a_2, a_3)$, we first observe   $
      \hat{U}^{(1)} (t)  =-\hat{U}^{(1)\dagger} (t), $ and $ 
       \hat{U} ^{(2)}(t)  =\hat{U}^{(2)\dagger} (t)$. 
Then  performing the integration, we can write        
       \begin{eqnarray}\label{eq}
           \text{Tr}_X[\hat{U}^{(1)} \hat{\rho} \hat{U} ^{(1)\dagger}]&=&-2\text{Tr}_X[\hat{U} ^{(2)}\hat{\rho} ]
       \end{eqnarray}
Now using these expressions, we observe  that $a_0 = a_1 =0$.  
To evaluate $a_2$, we observe that as $ \hat{H}(t)$ is a constraint, it commutes with every operator, including $\hat{U} $, and hence  we can write 
 \begin{eqnarray}
   a_2 &=&    \text{Tr}_X\Big[\hat{U}^{(1)} (t)   e^{-i \theta   \hat{H}(t)  }\hat{\rho} \hat{U} ^{(1)\dagger}(t)  e^{i \theta   \hat{H}(t)  }+\hat{U} ^{(2)}(t)   e^{-i\theta   \hat{H}(t)  }\hat{\rho} e^{i\theta   \hat{H}(t)  }   \nonumber\\ &&
      +e^{-i\theta   \hat{H}(t)  }\hat{\rho} e^{i\theta   \hat{H}(t)  }\hat{U} ^{(2)}(t)  \Big] =  0  \end{eqnarray}\\
       %Similarly, by repeating this calculation for the  the  coefficients of $|0\rangle\langle1|,|0\rangle\langle0|, $ and $|1\rangle\langle1|$, we observe that they also  vanish. 
       Similarly, we can demonstrate that $a_3 =0$. 
       Using  all these terms, we observe  
       \begin{equation}
        \hat{\rho}^2_\text{aux}(t) =0. 
       \end{equation}
So the qubit  state can be written as   $\hat{\rho}_\text{aux} (t) =|+\rangle\langle+|$   After perturbing the system by the dilaton field we now apply the second Hadamard gate, such that auxiliary qubit  is given by 
\begin{equation}
\hat{\rho}_\text{aux}(t )=|0\rangle\langle0| = \frac{1}{2}(\mathbb{1} +\hat{\sigma}_z). 
\end{equation}
It is known in general  \cite{6ab, 6ba}  
\begin{equation}
\label{20}
\hat{\rho}_\text{aux}    =
\frac12\Big[\mathbb{1} + \text{Re} [\tilde{P}(\theta)]\hat{\sigma}_z + \text{Im} [\tilde{P}(\theta)]\hat{\sigma}_y\Big]~. 
\end{equation}
Thus, we observe that the characteristic function is 
    \begin{equation}
    \tilde{P}(\theta)=1
 \end{equation} 
Using this characteristic function, we can calculate the average  difference between the initial and final eigenvalues of the Hamiltonian, and observe that it also vanishes
 \begin{eqnarray}
\langle \mathcal{H} \rangle  = -i \frac{d}{d\theta  }\tilde{P}(\theta  )|_{\theta  =0}=0, 
\end{eqnarray}
where $\mathcal{H}$ is the distribution variable, which is identified with quantum work  $\langle \mathcal{H} \rangle$.
So, the quantum work of the world-sheet Hamiltonian vanishes for such string theoretical processes.

\section*{Appendix 2. Mass-squared operator } 

Now we will probe the system by an alternative operator, i.e.,    
$ \hat{M}^2$.  Here we will still  perturb the system with 
a dilaton field, and evolve it by  $\hat{H}_I$.  Thus, the system still evolves by  the unitary operator, which was    defined in Eq. (\ref{a}), and then a Dyson expansion willl be  used to expand it. However, we  will  probe the effect of this perturbation using $\hat{M}^2$.

Thus, after   perturbing the system by the dilaton field,   we can write the unitary evolution of the system (string and auxiliary qubit) for $ \hat{M}^2$ as 
 \begin{eqnarray} \hat{C}_\theta(t) =\hat{U} (t)   e^{-i\theta \hat{M}^2(0) }\otimes|0\rangle\langle0|+ e^{-i\theta \hat{M}^2(t)   }\hat{U} (t)  \otimes|1\rangle\langle1|
 \end{eqnarray}
Now we can write the auxiliary qubit  at time $\tau=t$ as
    \begin{eqnarray}
  \hat{\rho}_\text{aux}(t)= \text{Tr} _{X}[\hat{C}_\theta(t)\hat{\rho}_\text{tot} \hat{C}_\theta^\dagger (t)]
    \end{eqnarray} 
 Using the Dyson expansion, we can again express this  reduced state of the qubit   as  
        \begin{equation}
              \hat{\rho}_\text{aux} (t) ={\hat{\rho}}^{0}_\text{aux} (t) +{\hat{\rho}}^{1}_\text{aux} (t)+{\hat{\rho}}^{2}_\text{aux} (t)+\mathcal{O}(\lambda^3)\end{equation}
             
 We can now evaluate these terms for the qubit. 
    The zeroth-order term is 
      \begin{eqnarray} 
      {\hat{\rho}}^{0}_\text{aux} (t) =|+\rangle\langle+|
    \end{eqnarray}
  The first-order term can be expressed as
     \begin{eqnarray}
   \hat{\rho}^1_\text{aux} (t) &=& \text{Tr}_X\Big[\big( \hat{U}^{(1)} (t)   e^{-i\theta   \hat{M}^2(0) }\otimes|0\rangle\langle0|  + e^{-i\theta   \hat{M}^2(t)  }\hat{U}^{(1)} (t)   \otimes|1\rangle\langle1|\big) \nonumber\\&&\times \hat{\rho}_\text{tot}\big(e^{i\theta   \hat{M}^2(0) }\otimes|0\rangle\langle0|+ e^{i\theta   \hat{M}^2(t)   }\otimes|1\rangle\langle1|\big) \Big] \nonumber  \\&&
    + \text{Tr}_X\Big[ \big(e^{-i\theta   \hat{M}^2(0) }\otimes|0\rangle\langle0|+ e^{-i\theta   \hat{M}^2(t)  } \otimes|1\rangle\langle1|\big)\hat{\rho}_\text{tot}\nonumber\\&&  \times \big(e^{i\theta   \hat{M}^2(0) }\hat{U}^{\dagger(1)}(t)  \otimes|0\rangle\langle0|   + \hat{U}^{\dagger(1)}(t)  e^{i\theta   \hat{M}^2(t)  }\otimes|1\rangle\langle1| \big)\Big]   \end{eqnarray}
The coefficient of $|0\rangle\langle0|$ term is  
      \begin{eqnarray}
      \text{Tr}_X[\hat{U}^{(1)}  e^{-i  \theta   \hat{M}^2}\hat{\rho}_\text{tot} e^{i  \theta   \hat{M}^2}+e^{-i  \theta   \hat{M}^2}\hat{\rho}_\text{tot} e^{i  \theta   \hat{M}^2}\hat{U}^{(1)\dagger} ] =0.
      \end{eqnarray} 
      Here we have again   used the cyclical property of trace  and  $\hat{U}^{(1)} =-\hat{U}^{(1)\dagger} $.  Using the same argument for all the coefficients in $ 
   \hat{\rho}^1_\text{aux}(t) $, we observe that they all vanish, and thus 
\begin{equation}
   \hat{\rho}^1_\text{aux}(t)   = 0 
\end{equation}
  Now we can write  $\hat{\rho}^2_\text{aux}(t)$  as
   \begin{eqnarray}\label{b}
   \hat{\rho}^2_\text{aux}(t)&=& 
 \text{Tr}_X\Big[\big(\hat{U}^{(1)} (t)   e^{-i\theta   \hat{M}^2(0) }\otimes|0\rangle\langle0| + e^{-i\theta   \hat{M}^2(t) }\hat{U}^{(1)} (t)  \otimes|1\rangle\langle1|\big)\nonumber \\ &&  \hat{\rho}_\text{tot}\big(e^{i\theta   \hat{M}^2(0) }\hat{U}^{(1)\dagger} (t) \otimes|0\rangle\langle0|  \hat{U}^{(1)\dagger} (t) e^{i\theta   \hat{M}^2(t)}\otimes|1\rangle\langle1|\big) \Big) \nonumber\\&&
+ \text{Tr}_X\Big(\big( \hat{U}^{(2)}(t)  e^{-i\theta   \hat{M}^2(0) }\otimes|0\rangle\langle0|   e^{-i\theta   \hat{M}^2(t) }\hat{U}^{(2)} (t)  \otimes|1\rangle\langle1|\big)\nonumber \\ &&  \hat{\rho}_\text{tot}\big(e^{i\theta   \hat{M}^2(0) }\otimes|0\rangle\langle0| e^{i\theta   \hat{M}^2(t)  }\otimes|1\rangle\langle1| \big)\Big] 
 \nonumber \\ && + \text{Tr}_X\Big[\big(e^{-i\theta   \hat{M}^2(0) }\otimes|0\rangle\langle0| + e^{-i\theta   \hat{M}^2(t) } \otimes|1\rangle\langle1|\big)\hat{\rho}_\text{tot} \nonumber \\ &&
\big(e^{i\theta   \hat{M}^2(0) }\hat{U}^{(2)\dagger} (t) \otimes|0\rangle\langle0|  +\hat{U}^{(2)\dagger } e^{i\theta   \hat{M}^2(t)  }\otimes|1\rangle\langle1| \big) \Big]
   \end{eqnarray}
 This expression is again of the form 
\begin{equation}
  \hat{\rho}^2_\text{aux}(t) = a_0 \otimes|0\rangle\langle0|+ a_1 \otimes|1\rangle\langle1| + a_2 \otimes|1\rangle\langle0|+ a_3 \otimes|0\rangle\langle1|
\end{equation}
Here we can write $a_2 = a_{21} + a_{22}$, where $a_{21}$ is the contribution from $U^{(1)}$ and $U^{(1)\dagger}$ and $a_{22}$ is the contribution from $U^{(2)}$ and $U^{(2)\dagger}$. %To evaluate these terms, we first note  that using the cyclic property of trace,  we can write  
  % \begin{eqnarray}
   %   \hat{U}^{(1)} (t) =-\hat{U}^{(1)\dagger} (t) , && 
    %           \hat{U} ^{(2)}(t) =\hat{U}^{(2)\dagger} (t) ,  \nonumber\\
     %  e^{-i \theta   \hat{M}^2}\hat{\rho} e^{i \theta   \hat{M}^2}=\hat{\rho} && \end{eqnarray}
      % We can also write $\text{Tr}_X[\hat{U}^{(1)} \hat{\rho} \hat{U} ^{\dagger(1)}]$ as   
       %\begin{eqnarray}
        %   \text{Tr}_X[\hat{U}^{(1)} \hat{\rho} \hat{U} ^{
         %  \dagger(1)}]=-2\text{Tr}_X[\hat{U}^{(2)}  \hat{\rho}]
      % \end{eqnarray}
      Using Eq. (\ref{eq}), 
 we observe that   coefficients of diagonal  terms    $|0\rangle\langle0|$ and $|1\rangle\langle1|$  vanish, i.e., $a_0=a_1 =0$.  
Using  the property of  $\hat{M}^2$ operator, i.e., the eigenvalue of $\hat{M}^2$ operator  on the $n$ excited state is  ${(n-1)}/{\alpha'}$, the coefficient $a_{21}$ can be expressed as  
    \begin{eqnarray} a_{2 1} &=&
\frac{1}{2} \text{Tr}_X\Big[\hat{U}^{(1)} (t) e^{-i \theta \hat{M}^2}\hat{\rho} \hat{U}^{(1)\dagger} (t) e^{i \theta \hat{M}^2}\Big]  \nonumber \\ 
  &=&\mathcal{N} \text{Tr}_X\big[\lambda^2 \alpha'\sum_{n\neq0}\sum_{m\neq0}\frac{1}{nm}\int d\tau \chi(\tau)\int d\tau'\chi(\tau')\int  R c_\mu  \cos(n\sigma)d\sigma \nonumber \\
 &&  \times
\int   Rc^\nu d{\sigma'}e^{i n\tau} \sum_k\sum_l d_k d_l 
e^{-i \theta \big(\frac{n+k-1}{\alpha'}\big)}(\hat{\alpha}^\mu  _{-n}|k;p\rangle\langle l;p|\hat{ \alpha}_{\nu m} ) \nonumber\\ && e^{i \theta  \big(\frac{l-1}{\alpha'}\big)} e^{-i  m\tau'} \cos(m\sigma')\Big]
  \end{eqnarray}
Now observing that    $
\langle k;p|\hat{\alpha}_{\nu{n}}\hat{\alpha}^\mu _{-n}|l;p\rangle=\delta^\mu _\nu\delta^k_l\sqrt{n+l}\sqrt{n+k}
        $, we can write 
       % We can write  the  contribution to  coefficient of  the $|1\rangle\langle0|$  from terms containing  $\hat{U} ^{(1)}..\hat{U} ^{(1)\dagger}$ in Eq. (\ref{b}) \jam{equation 52 doesnt exists} as
       \begin{eqnarray}\label{c} 
         a_{21} =     \mathcal{N} \lambda^2 \alpha'\left[{\sum^\infty_{n=1}\sum_{k=0}^\infty \frac{d^2_k}{n^2}(n+2k)|\Tilde{\chi}(n)|^2\Phi(n)^{\mu }\Phi_{\mu  } (n)e^{-i \frac{\theta  }{\alpha'}n}}\right]
        \end{eqnarray}
    Here $\Tilde{\chi}(n)$ is a Fourier transformation  of $\chi(\tau)$ and $|\Phi(n)|$ is the cosine transformation of $R c_\mu  $. %Coefficient of  the $|0\rangle\langle1|$  from terms containing  $\hat{U} ^{(1)}...\hat{U} ^{(1)\dagger}$ will be Hermitian conjugate of the  Eq. (\ref{c}) \jam{equation doesnt exists}. 
    The  coefficient  $a_{22}$ can be expressed as 
    \begin{equation}
      a_{22} =    - \mathcal{N} \lambda^2 \alpha'\left[{\sum^\infty_{n=1} \sum_{k=0}^\infty \frac{d^2_k}{n^2}(n+2k)|\Tilde{\chi}(n)|^2\Phi(n)^{\mu }\Phi_{\mu  } (n)}\right]
        \end{equation}
We can also write  the coefficient $a_3 = a_{31} + a_{32}$, with $a_{31}$ being the coefficient of   $|0\rangle\langle1|$  from $U^{(1)}$ and $U^{(1)\dagger}$, and $a_{31}$ is the coefficient of 
 $|1\rangle\langle0|$  from   $U^{(2)}$ and $U^{(2)\dagger}$. Now we observe that $a_{31}$ is the complex conjugate of $a_{21}$, $a_{31} = \bar{a}_{21}$, and $a_{32} = a_{22}$.  
Using these expression, the    qubit can be write  as   
   \begin{eqnarray}
  \hat{\rho}_\text{aux}(t) &=&  \mathcal{N} \lambda^2 \alpha'\left[{\sum^\infty_{n=1}\sum_{k=0}^\infty \frac{d^2_k}{n^2}(n+2k)|\Tilde{\chi}(n)|^2\Phi(n)^{\mu }\Phi_{\mu  } (n)\big(e^{-i \frac{\theta  }{\alpha'}n}-1\big)}\right]\frac{1}{2}|1\rangle\langle0|  \nonumber\\  &&
 +\mathcal{N} \lambda^2 \alpha'\left[{\sum^\infty_{n=1}\sum_{k=0}^\infty \frac{d^2_k}{n^2}(n+2k)|\Tilde{\chi}(n)|^2\Phi(n)^{\mu }\Phi_{\mu  } (n)\big(e^{i \frac{\theta  }{\alpha'}n}}-1\big)\right]\frac{1}{2}|0\rangle\langle1|  \nonumber\\
 && +\, \,  |+\rangle\langle+|  
   \end{eqnarray}
Now we apply the second Hadamard, to obtain the final state of the 
 qubit. Using this state of the qubit, and Eq.  (\ref{20}), 
we  can  explicitly  write the  characteristic function as 
  \begin{eqnarray}
\tilde{P}(\theta)&=& 
 \mathcal{N}  \lambda^2 \alpha'\sum^\infty_{n=1}\sum_{k=0}^\infty \frac{d^2_k}{n^2}(n+2k)|\Tilde{\chi}(n)|^2\Phi(n)^{\mu }\Phi_{\mu  }(n)\Big({\cos\left(\frac{\theta   n}{\alpha'}\right)}-1\Big)
 \nonumber \\&&
+i\mathcal{N}  \lambda^2 \alpha'\sum^\infty_{n=1}\sum_{k=0}^\infty \frac{d^2_k}{n^2}(n+2k)|\Tilde{\chi}(n)|^2\Phi(n)^{\mu }\Phi_{\mu  }(n)\left({\sin\left(\frac{\theta   n}{\alpha'}\right)}\right)\nonumber \\&&  
+\, \, 1 
  \end{eqnarray} 
We can write the $\hat{M}^2$ analog for quantum work as 
  \begin{equation}
  \langle \mathcal{M}^2\rangle=-i \frac{d}{d\theta  }\tilde{P}(\theta  )|_{\theta  =0}, 
\end{equation}
where $\mathcal{M}^2$ is the distribution variable. 
This measures the  average of the difference between the initial and final eigenvalues of $M^2$, and can be explicitly written  as   
\begin{equation}
      \langle \mathcal{M}^2\rangle=\mathcal{N} \lambda^2 \sum^\infty_{n=1}\sum_{k=0}^\infty \frac{d^2_k}{n}(n+2k)|\Tilde{\chi}(n)|^2\Phi(n)^{\mu }\Phi_{\mu  }(n).
\end{equation}
Now we observe that this does not generally vanish, as $|\Tilde{\chi}(n)|^2 \neq 0 $ and $\Phi(n)^{\mu }\Phi_{\mu  }(n) \neq 0$. 

\end{document}